\documentstyle[aps,preprint,epsfig]{revtex}

\begin{document}
\def\bea{\begin{eqnarray}}
\def\eea{\end{eqnarray}}
\def\nn{\nonumber}
\renewcommand\epsilon{\varepsilon}
\def\beq{\begin{equation}}
\def\eeq{\end{equation}}
\def\lla{\left\langle}
\def\rra{\right\rangle}
\def\za{\alpha}
\def\zb{\beta}
\def\lsim{\mathrel{\raise.3ex\hbox{$<$\kern-.75em\lower1ex\hbox{$\sim$}}} }
\def\gsim{\mathrel{\raise.3ex\hbox{$>$\kern-.75em\lower1ex\hbox{$\sim$}}} }
\newcommand{\Rbs}{\mbox{${{\scriptstyle \not}{\scriptscriptstyle R}}$}}

\draft

\title{New Interpretation on the Solar Neutrino Flux \\
with Flavor Mixing and Majorana Magnetic Moment
}

\author{ S.~ K.~ Kang$^{a}$\thanks{ E-mail : skkang@phya.snu.ac.kr}~ and~
C.~ S.~ Kim$^{b}$ \thanks{E-mail : cskim@yonsei.ac.kr}}
\date{\today}
\address{\small \it $^{a}$School of Physics, Seoul National
University,
       Seoul 151-734,Korea \\
       $^{b}$Department of Physics, Yonsei University, Seoul 120-749, Korea}
\maketitle \thispagestyle{empty}
\begin{abstract}
A simple and model-independent method is proposed to extract
information on $\nu_e$ transition into antineutrinos via the spin
flavor precession (SFP) from the measurements of solar neutrino
flux at SNO and Super-Kamiokande. Incorporating the KamLAND
experimental results, we examine how large the suppression of the
solar neutrino flux could be due to the SFP mechanism in the
context of the hybrid scenario with two-flavor neutrino and
antineutrino mixings.

\end{abstract}
\widetext
Thanks to the recent neutrino experiments at Sudbury Neutrino
Observatory (SNO) \cite{SNOCC,SNONC} and KamLAND  \cite{kamex},
the resolution of the long-standing solar neutrino problem,
discrepancy between the prediction of the neutrino flux based on
the standard solar model (SSM) \cite{ssm} and that measured by
experiments, is just around corner. In addition to the
radiochemical experiments on $Ga$  and $Cl$ targets for solar
neutrinos, the water Cerenkov experiments from Super-Kamiokande
(SK) \cite{SK2002} has observed the emitted electron from elastic
scattering ($ES$) $ \nu_{x} + e \rightarrow \nu_{x} + e,
\label{one} $ ($\nu_x=\nu_e,\nu_{\mu},\nu_{\tau}$), while SNO has
measured the neutrino flux  through the charged current ($CC$)
process $\nu_e + d \rightarrow p + p + e \label{two},$ the neutral
current ($NC$) process $ \nu + d \rightarrow \nu + p + n,
\label{three} $ and $ES$ process given in the above. Both the
experiments, SNO and SK,  probe the high energy tail of the solar
neutrino spectrum, which is dominated by the $^8B$ neutrino flux.
The results of the solar neutrino flux measured at SK and SNO are
as follows \cite{SNONC,SK2002}:
\begin{eqnarray}
  \label{eq:8Bexprates}
\label{eq:CC}
  \Phi_{\rm SNO}^{CC} &=&
  1.76\pm 0.11\times 10^6 \mbox{cm}^{-2}\mbox{s}^{-1},\\
\label{eq:ESSNO}
  \Phi_{\rm SNO}^{ES} &=&
  2.39\pm 0.27\times 10^6 \mbox{cm}^{-2}\mbox{s}^{-1},\\
\label{eq:ESSK}
  \Phi_{\rm SK}^{ES} &=&
  2.35\pm 0.08\times 10^6 \mbox{cm}^{-2}\mbox{s}^{-1},\\
\label{eq:NC1}
  \Phi_{\rm SNO}^{NC1} &=&
  5.09\pm 0.63\times 10^6 \mbox{cm}^{-2}\mbox{s}^{-1},\\
\label{eq:NC2}
  \Phi_{\rm SNO}^{NC2} &=&
  6.42\pm 1.67\times 10^6 \mbox{cm}^{-2}\mbox{s}^{-1},
\end{eqnarray}
where $\Phi_{\rm SNO}^{NC1}$ was found
  assuming undistorted $^8B$ neutrino energy spectrum, while $\Phi_{\rm
    SNO}^{NC2}$ was found when this assumption was relaxed.
Comparing those results with the prediction of the SSM,
$\Phi_{\rm SSM}=5.05^{+1.01}_{-0.81} \times 10^6 \mbox{cm}^{-2}\mbox{s}^{-1}$
\cite{ssm}, we confirm the flux deficit of solar neutrino. It is
commonly believed that the most plausible solution of the solar
neutrino anomaly is in terms of neutrino oscillation, and in
particular the oscillation of $\nu_e$ into another active flavor
$\nu_a$, which can be any combination of $\nu_\mu$ and $\nu_\tau$.
Based on a global analysis in the framework of two-active neutrino
oscillations of all solar neutrino data, the large mixing
angle (LMA) solution is favored and oscillations into a pure
sterile state are excluded at high confidence level \cite{lma}. On
the other hand, the KamLAND collaboration \cite{kamex} has found
for the first time the evidence for the disappearance of reactor
antineutrinos, which confirms LMA solution of the solar neutrino
problem for the CPT invariant neutrino mass spectrum. It appears
that all non-oscillation solutions of the solar neutrino problem
are strongly disfavored \cite{kamth}.

The spin flavor precession (SFP) solution of the solar neutrino
problem \cite{SFP}, motivated by the possible existence of nonzero
magnetic moments of neutrinos, has also been studied and  found to
have a good fit slightly better than the LMA oscillation solution
before the KamLAND experiment.  Although the KamLAND result
excludes a {\it pure} SFP solution to the solar neutrino problem
under the CPT invariance, a fraction of the flux suppression of
solar neutrino may still be attributed to SFP \cite{comb1,comb2}.
In this respect, we believe that the detailed investigation on how
much the flux suppression of solar neutrino can be attributed to
SFP will lead us to make considerable progress in understanding
the solar neutrino anomaly as well as the inner structure of the
Sun. In this Letter, we propose a simple and model-independent
method to extract information on $\nu_e$ transition into
antineutrinos via SFP from the measurements of $^8B$ neutrino flux
at SNO and SK. As will be seen, in particular, our determination
of the mixing between non-electron active neutrino and
antineutrino is not affected by the existence of transition into a
sterile state. In addition, we  study the implication of KamLAND
results on the hybrid scenario with both neutrino oscillation and
SFP conversion.

Let us begin by considering how the experimental measurement of
the solar neutrino flux can be presented in terms of the solar
neutrino survival probability. In this study, we use $ ES$
measurement of SK instead of corresponding measurement of SNO, and
take $ \Phi_{\rm SNO}^{NC1}$ as $NC$ flux because of a smaller
error. Assuming the SSM neutrino fluxes and the transition of
$\nu_e$ into a mixture of  active flavor $\nu_a$ and sterile
$\nu_s$, $\nu_a \sin\alpha + \nu_s \cos\alpha$, that participate
in the solar neutrino oscillations, one can write the SK $ES$ rate
and the SNO $CC$ and $NC$ scattering rates relative to the SSM
predictions in terms of the survival probability
\cite{barger,snofit}:
\begin{eqnarray}
R^{ES}_{\rm SK} &\equiv& \Phi^{ES}_{\rm SK} /\Phi_{\rm SSM} =
                f_B [P_{ee} + r \sin^2\alpha (1-P_{ee})],
\label{six} \\[2mm]
R^{CC}_{\rm SNO} &\equiv & \Phi^{CC}_{\rm SNO} /\Phi_{\rm SSM}
               = f_B P_{ee},
\label{seven} \\[2mm]
R^{NC}_{\rm SNO} &\equiv & \Phi^{NC}_{\rm SNO} /\Phi_{\rm SSM}
                = f_B [P_{ee} +\sin^2\alpha (1-P_{ee}) ],
\label{eight}
\end{eqnarray}
 where
$r =\sigma^{NC}_{\nu_a}/\sigma^{CC+NC}_{\nu_e} \simeq 0.171$
   is the
ratio of $\nu_{\mu,\tau}$ to $\nu_e$ $ES$ cross-sections
\cite{bahcall}, and $P_{ee}$ is the $\nu_e$ survival probability.
Here $\sin^2\alpha$ indicates the fraction of
 $\nu_e$ oscillation to active flavor $\nu_a$.
Since there is a large uncertainty in the predicted
normalization, $\Phi_{\rm SSM}$,
arising from the uncertainty in the
$^7Be + p \rightarrow \ {^8B} + \gamma $
cross-section, we have introduced a
constant parameter $f_B$ to denote the normalization of the $^8B$
neutrino flux relative to the SSM prediction.
It should be also noted here that
the SK $ES$  and SNO $CC$ data start from neutrino energies
of 5 and 7 MeV respectively, while the response function of the
SNO $NC$ measurement extends marginally below 5 MeV. However, the
SK rate and the resulting survival probability show energy
independence down to 5 MeV to a very good precision. The SNO
$CC$ rate shows energy independence as well, although to lesser
precision. Therefore, it is reasonable to assume a common survival
probability for all the three measurements.
Using the measured values of the rates $R$, we can estimate the
allowed regions of the quantities $f_B, P_{ee}$ and $\alpha$. In
particular, the fraction of $\nu_e$ oscillation to $\nu_a$ is
described by the relation,
\begin{equation}
\sin^2\alpha = \frac{R_{\rm SK}^{ES}-R_{\rm SNO}^{CC}}
{r(f_B-R^{CC}_{\rm SNO})}. \label{rel1}
\end{equation}
As is well known, a family of the present solar neutrino
experiments is not enough to extract the value of $\alpha$ because
of the unknown parameter $f_B$. This is so-called the $\alpha,
f_B$ degeneracy \cite{barger}. The recent analysis shows that a
pure sterile oscillation solution $(\sin^2 \alpha=0)$ is
disfavored and $0.14\leq \sin^2 \alpha \leq 1$ is obtained for
$f_B\leq 2$ from the $\chi^2$ analysis at the $1\sigma$
level\cite{barger}.

{\bf Model independent analysis for SFP + Oscillation :}\\
So far, we
have considered only the possible $\nu_e$ conversion to the active
and sterile neutrinos. The excess $NC$ and $ES$ events can also be
caused by the antineutrinos.
 The antineutrinos in question
must be of the muon or tau types, since $\bar\nu_e$ would be
easily identified through the reaction $\bar\nu_e+p\rightarrow
n+e^+$ \cite{antiel}.
 Both $\nu_{\mu,\tau}$ and
$\bar\nu_{\mu,\tau}$ scatter on electrons and deuterium nuclei
through their $NC$ interactions, with different cross sections.
The $\nu_e\rightarrow\nu_{\mu,\tau}$ conversion is predicted in
the flavor oscillation scenario, while
$\nu_e\rightarrow\bar\nu_{\mu,\tau}$ is predicted in the neutrino
SFP scenario \cite{SFP}. When we allow the possible conversion to
the antineutrinos, the previous formula (\ref{six}) for the SK $ES$ rate
relative to the SSM prediction is modified as
follows:
 \bea R^{ES}_{\rm SK} &=& f_B (P_{ee} +
r\sin^2\alpha
\sin^2 \psi (1-P_{ee}) \nonumber \\
& &+ \bar{r}\sin^2 \alpha \cos^2 \psi (1-P_{ee})), \label{rel2}
\label{twelve-cc} \eea where
$\bar{r}=\sigma^{NC}_{\bar{\nu}_a}/\sigma^{CC+NC}_{\nu_e} \simeq
0.114$ and $\psi$ is a mixing angle that describes the linear
combination of the probability of $\nu_e$ conversion into $\nu_a$
and active antineutrinos.
Assuming the conversions between two families, we have
$\tan^2 \psi=P(\nu_e\rightarrow \nu_{\mu})/P(\nu_e\rightarrow
\bar{\nu}_{\mu})$.

{}From Eqs. (\ref{seven},\ref{eight},\ref{rel2}), we see that the
mixing angle $\psi$ is related with the measured neutrino fluxes
as follows:
\begin{eqnarray}
r\sin^2\psi + \bar{r}\cos^2\psi =
 \frac{R^{ES}_{\rm
SK}-R^{CC}_{\rm SNO}}{R^{NC}_{\rm SNO}-R^{CC}_{\rm SNO}} ,
\label{rel3}
\end{eqnarray}
where we have assumed that $\sin^2\alpha$ is non-zero. The
expression (\ref{rel3}) shows that  the determination of the
mixing angle $\psi$ is independent of $\sin^2\alpha$. While the
mixing angle $\alpha$ could not be determined from the SNO and SK
data due to the $\alpha, f_B$ degeneracy, it appears that we are
led to determine the mixing angle $\psi$ from those data. As one
can see from Eq. (\ref{rel3}), the precise measurements of
$R^{ES}_{\rm SK},R^{NC}_{\rm SNO},R^{CC}_{\rm SNO}$ as well as the
values of $r$ and $\bar{r}$ make it possible to see how much the
solar neutrino flux deficit can be caused by SFP. We note that any
deviation of the value of $\sin^2\psi$ from one implies the
evidence for the existence of $\nu_{e}$ transition into
non-sterile antineutrinos, and if there is no transition of solar
neutrino due to the magnetic field inside the sun, the left-hand
side of Eq. (\ref{rel3}) should be identical to the parameter $r$.
Using the experimental data, we obtain the values for the
right-hand side of Eq. (\ref{rel3}):
\begin{eqnarray}
0.169\pm 0.053 . \label{num}
\end{eqnarray}
We see that the central value is close to the value of $r$ for the
energy threshold of SNO and SK, which implies that the best fit
corresponds to the very small possibility of $\nu_e \rightarrow
\nu_{\bar{a}}$ transition even if large possibility of such a
transition is allowed. Since $\bar{r} \leq r\sin^2\psi +
\bar{r}\cos^2\psi \leq r$, we notice that the left-hand side of
Eq. (\ref{rel3}) prefers to lower side of Eq. (\ref{num}).  The
central value of Eq. (\ref{num}) leads to $\sin^2\psi=0.94$. Since
the error in Eq. (\ref{num}) which leads to $\delta
\sin^2\psi=0.93$ is rather large, we cannot obtain any severe
constraint on the mixing angle $\psi$ from the present solar
neutrino experimental results, but we can say that the spin-flavor
transition due to neutrino magnetic moment is allowed in the light
of the high energy solar neutrino experiments, SNO and SK. In the
future, experiments may lead the error in Eq. (\ref{num}) to be
reduced so that a lower bound on the mixing angle $\psi$ could be
obtained. For example, if the future SNO experiments could reduce
the error in charged current measurement to $50\%$, then the
uncertainty on $\sin^2\psi$ becomes $0.76$, and if the errors in
both charged and neutral current experiments could be reduced to
$50\%$, the uncertainty on $\sin^2\psi$ becomes $0.60$. Therefore,
there is room for resolving the solar neutrino flux deficit
through both neutrino oscillation and SFP.

{\bf Implication of KamLAND} : \\
The recent reactor neutrino measurement at KamLAND implies the
existence of  $\nu_e$ oscillation in vacuum. The allowed regions
for the vacuum mixing angle and $\Delta m^2$ at $3\sigma$ are
given by \cite{kamth}
\begin{eqnarray}
0.29 &\leq & \tan^2\theta \leq 0.86, \nonumber \\
5.1\times 10^{-5} \mbox{eV}^2 &\leq & \Delta m^2 \leq 9.7\times
10^{-5} \mbox{eV}^2, \nonumber \\
1.2\times 10^{-4} \mbox{eV}^2 &\leq & \Delta m^2 \leq 1.9\times
10^{-4} \mbox{eV}^2. \label{kam}
\end{eqnarray}
The local minimum occurs for $\tan^2\theta=0.42, \Delta m^2 =
1.4\times 10^{-4} \mbox{eV}^2$. In the light of KamLAND
experimental result, the large mixing angle MSW solution of the
solar neutrino problem is strongly favored \cite{kamth}.
Let us consider the implication of the
KamLAND result on the solar neutrino problem when we allow both
neutrino oscillation and SFP mechanism as discussed in the
previous section. As will be shown later, in order to determine
both neutrino conversion effects, we need to know the mixing angle
and $\Delta m^2$ in vacuum as well as the information on the Sun
such as matter density and magnetic field profile inside the Sun.
In this Letter, we will take the KamLAND results to fix the mixing
angle and $\Delta m^2$ in vacuum and then investigate how the
parameters concerned with the Sun can be constrained by the SNO
and SK experiments.

Assuming two Majorana families, $\nu_e$ and $\nu_{\mu}$,
 without sterile sectors,
the evolution equations for the two families  can be presented in terms of
$4\times 4$ Hamiltonian matrix in the basis $(\nu_e, \nu_{\mu},
\bar{\nu}_e,\bar{\nu}_{\mu})$ in a nonzero transverse magnetic
field as given in Ref. \cite{SFP2}
\small
\begin{eqnarray}
H=\left( \begin{array}{cccc}
a_{\nu_e} & \frac{\Delta m^2}{4E_{\nu}}\sin 2\theta & 0 & \mu^{*}_{\nu} B \\
\frac{\Delta m^2}{4E_{\nu}}\sin 2\theta & \frac{\Delta
m^2}{2E_{\nu}}\cos 2\theta +  a_{\nu_{\mu}} &
-\mu^{*}_{\nu} B & 0 \\
0 & -\mu_{\nu} B & -a_{\nu_e} & \frac{\Delta m^2}{4E_{\nu}}\sin
2\theta \\
\mu_{\nu} B & 0 & \frac{\Delta m^2}{4E_{\nu}}\sin 2\theta &
\frac{\Delta m^2}{2E_{\nu}}\cos 2\theta -  a_{\nu_{\mu}}
\end{array} \right )
\end{eqnarray}
\normalsize
 where $\mu_{\nu}$ is neutrino magnetic moment and $B$
is magnetic field strength inside the Sun and
$a_{\nu_e}=G_{\mu}/\sqrt{2}(2N_e-N_n),
a_{\nu_{\mu}}=G_{\mu}/\sqrt{2}(-N_n)$ where $N_e$ and $N_n$ are
the densities of electron and neutrons in the Sun.
 This Hamiltonian
can be diagonalized exactly by
\begin{eqnarray}
X=\left( \begin{array}{cccc}
  c^a & s^a & 0 & 0 \\
  -s^a & c^a & 0 & 0 \\
  0 & 0 & c^b & s^b \\
  0 & 0 & -s^b & c^b \end{array} \right)
  \left( \begin{array}{cccc}
  1 & 0 & 0 & 0 \\
  0 & c^{\prime} & s^{\prime} & 0 \\
  0 & -s^{\prime} & c^{\prime} & 0 \\
  0 & 0 & 0 & 1 \end{array}
  \right)
  \left( \begin{array}{cccc}
   c^{\prime \prime}& 0 & 0 & s^{\prime \prime} \\
   0 & 1 & 0 & 0 \\
   0 & 0 & 1 & 0 \\
   -s^{\prime \prime} & 0 & 0 & c^{\prime \prime} \end{array}
   \right)
   \end{eqnarray}
   where $s^a=\sin \theta^a, s^{\prime}=\sin \theta^{\prime}$ and
   $s^{\prime \prime} = \sin \theta^{\prime \prime}$ and
   $c^a,c^{\prime},c^{\prime \prime}$ correspond to the cosines of
   the mixing angles.
For simplicity, let us consider the adiabatic approximation which
implies slowly changing magnetic field and matter density inside
the Sun. Then, without the oscillating term, transition
probabilities are simply given as follows:
\begin{eqnarray}
P(\nu_e\rightarrow \nu_e) &=&
             \cos^2 \theta^{\prime \prime}_p(\cos^2\theta \cos^2 \theta^a_p
             +\sin^2\theta \sin^2 \theta^a_p), \label{pee} \\
P(\nu_e\rightarrow \nu_{\mu}) &=&
             \cos^2 \theta^{\prime}_p(\cos^2\theta \sin^2 \theta^a_p
             +\sin^2 \theta \cos^2 \theta^a_p), \\
P(\nu_e\rightarrow \bar{\nu}_{\mu}) &=&
             \sin^2 \theta^{\prime \prime}_p(\cos^2\theta \cos^2 \theta^a_p
             +\sin^2\theta \sin^2 \theta^a_p), \\
P(\nu_e\rightarrow \bar{\nu}_{e}) &=&
         \sin^2 \theta^{\prime}_p(\cos^2\theta \sin^2 \theta^a_p
            +\sin^2 \theta \cos^2 \theta^a_p),
\end{eqnarray}
where $\theta$ is the mixing angle in vacuum, whereas
$\theta^{\prime}_p, \theta^{\prime \prime}_p$ and $\theta^a_p$ are
the mixing angles at the production point of $\nu_e$ inside the
Sun. Note that the mixing angles $\theta^{\prime}_p$ and
$\theta^{\prime \prime}_p$ depend on $ \mu B, \theta, \Delta m^2/
E_{\nu}$ as well as the matter densities $a_{\nu_e}$ and
$a_{\nu_{\mu}}$. Non-adiabatic case as well as the details on the
mixing matrix $X$ and the exact expressions of the probabilities
will be studied elsewhere \cite{skk}. When neutrinos are produced
in the region where the matter density is far above the resonance
region $(a_{\nu_e}-a_{\nu_{\mu}} =\Delta m^2/(2E_{\nu}))$,
$\theta^a_p$ is close to $\pi/2$ at the production point as in the
case of MSW solution. We also note that there are no resonance
regions for $\theta^{\prime}_p$ and $\theta^{\prime \prime}_p$
\cite{skk}.

 The bound on the probability $P(\nu_e\rightarrow
\bar{\nu}_{e})\leq 1.5\%$ \cite{comb1} may lead us to see how
large are the values of parameters concerned with solar magnetic field such as
$\mu B$, $\theta^{\prime}_{p}$ and $\theta^{\prime \prime}_{p}$.
For given values of $\theta^a_p$ and $\theta$, we can obtain
some bound on $\theta^{\prime}_p$ which in turn leads us to
bounds on the parameters such as $\mu B, \theta^{\prime \prime}$
and $\psi$. In particular, in the limit of $\theta^a_p=\pi/2$, we
note that the relations, $(\mu B)^2 = a^2_{\nu_e}
\tan^2{2\theta^{\prime}_p}$, holds and the mixing angle $\psi$ can
be presented in terms of $\theta,\theta^{\prime}_p$, and
$\theta^{\prime\prime}_p$,
$\sin^2\psi \simeq 1-\sin^2\theta^{\prime \prime}_p/\tan^2\theta
\cos^2\theta^{\prime}_p $. In our numerical analysis, we consider
several cases with fixed values of $\sin^2\theta_p^{\prime}$ which
correspond to the current bound and soon achievable sensitivities
on $P(\nu_e\rightarrow \bar{\nu}_{e})$. Imposing the KamLAND
results given in Eq.(\ref{kam}) and $\theta^a_p\sim \pi/2$, we can
estimate the allowed regions of the parameters $\theta^{\prime
\prime}_p$ and $\psi$. This procedure makes us to predict the
possible values of the survival probability $P(\nu_e\rightarrow
\nu_{e})$. We have also found that the numerical results are not
sensitive to $^8 B$ neutrino energies $E_{\nu}$ measured in SK and
SNO. In Fig. 1, we plot $P(\nu_e\rightarrow \nu_{e}) ~vs.~ \mu B$
for $P_{e\bar{e}}= $ (a) 0.015, (b) 0.005, (c) 0.001 and
(d) 0.0001. The solid lines show how the value of
$P(\nu_e\rightarrow \nu_{e})$ depends on $\mu B$ for the allowed
ranges of $\theta$, $\Delta m^2/E_{\nu}$ and fixed values of
$P_{e\bar{e}}$. The two vertical lines in Fig. 1 are the upper and
lower limits on the allowed region of $P(\nu_e \rightarrow\nu_e)$
at $1\sigma$ from the solar neutrino experimental results,
$P(\nu_e \rightarrow\nu_e)=0.35\pm 0.07$  assuming $f_B=1$
\cite{barger,snofit}.
In Fig. 2, we plot $P(\nu_e\rightarrow
\nu_{e})~ vs.~ \sin^2\psi$ for the same fixed values of
$P(\nu_e\rightarrow \bar{\nu}_{e})$ as in Fig. 1. The two vertical
lines are also the same as in Fig. 1. Here, we note that any
deviation of $\sin^2\psi$ from one implies the existence of solar
$\nu_e$ transition into active antineutrinos via SFP mechanism. As
one can see from Fig. 2, there is some region of $\sin^2\psi$
deviated from one which is consistent with the current solar
neutrino observations and KamLAND experiment. We can obtain a
bound on $\sin^2\psi\sim 0.76$ which corresponds to $P(\nu_e
\rightarrow \nu_e)\sim 0.28$ and the case of $P(\nu_e\rightarrow
\bar{\nu}_e)=0.015$. As anticipated, $\sin^2\psi$ approaches to
one as $P(\nu_e\rightarrow \bar{\nu}_e)$ becomes smaller.
From the results, we see that some part of the
suppression of solar neutrino flux may be due to the conversion of
$\nu_e$ to $\bar{\nu}_{\mu}$ via SFP mechanism although it is not
without any uncertainty.
Thus, we conclude that the hybrid scenario with both
neutrino oscillation and SFP conversion may be presently consistent with
solar neutrino experiments and KamLAND result.

In summary, we have proposed a simple and model-independent method
to extract information on $\nu_e$ transition into antineutrinos
via SFP from the measurements of solar neutrino flux at SNO and
SK. But, from the current solar neutrino experiments, we could not
obtain any severe constraint as to how large $\nu_e$ transition
into antineutrinos via SFP could be. Incorporating the KamLAND
experimental results, we have examined how large the suppression
of solar neutrino flux could be due to the SFP mechanism in the
context of the hybrid scenario with two-flavor neutrino and
antineutrino mixings.
\\

S.K.K is supported by and by BK21 program of the Ministry of
Education in Korea, and by Korea Research Foundation Grant
(KRF-2002-015-CP0060). The work of C.S.K. was supported by Grant
No. R02-2003-000-10050-0 from BRP of the KOSEF.

\begin{figure}
\begin{center}
    \leavevmode
    \epsfysize=8.0cm
    \epsffile[75 160 575 530]{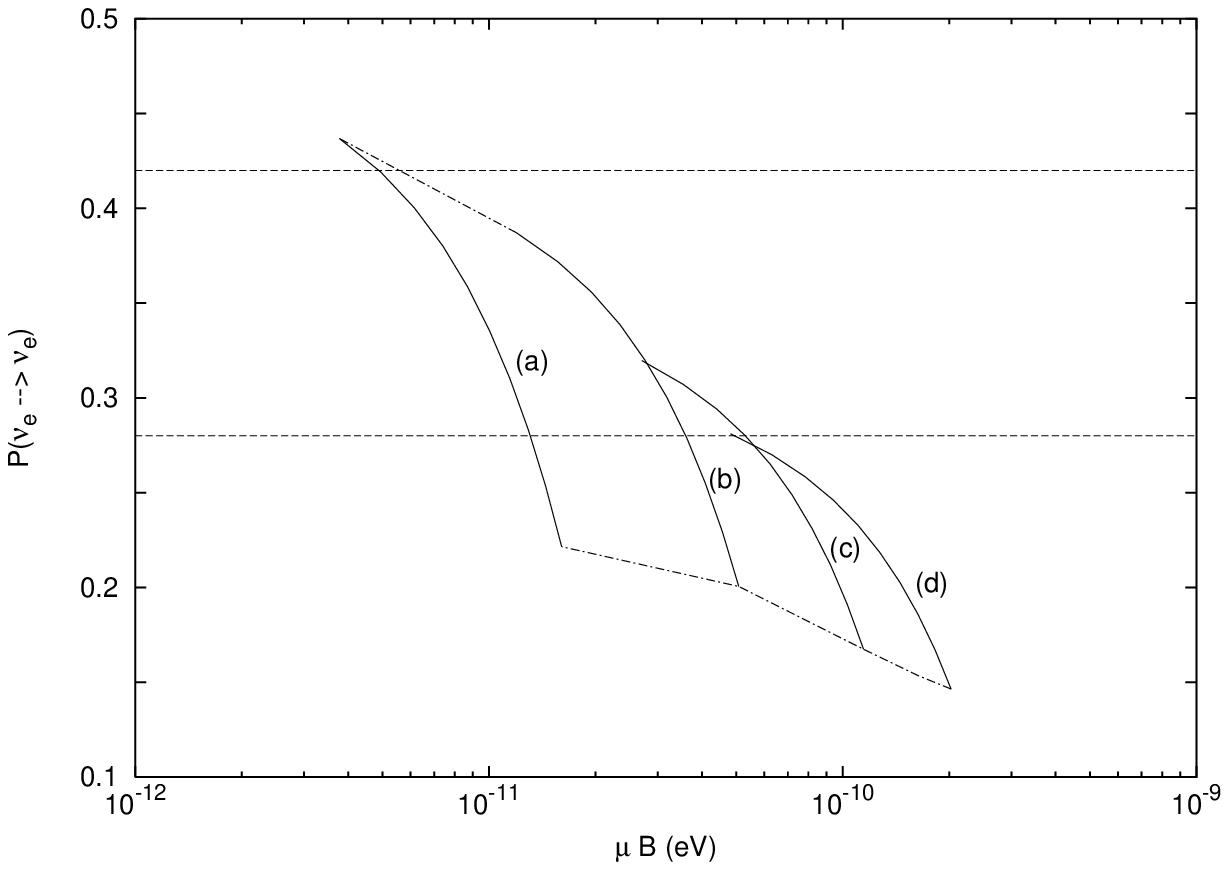}
   \vspace{5.0cm}
\smallskip
\caption{\label{f1} The prediction of $P(\nu_e\rightarrow \nu_e)$
as a function of
 $\mu B$ for the ranges of $ \tan^2 \theta$ and $\Delta m^2$ given in Eq.
(\ref{kam}) and $P(\nu_e\rightarrow \bar\nu_e)$= (a) 0.015, (b) 0.005,
(c) 0.001 and (d) 0.0001. The two vertical lines correspond
to the upper and lower limits on $P(\nu_e\rightarrow \nu_e)$ at
$1\sigma$ from the solar neutrino experimental results. }
\end{center}
\end{figure}

\begin{figure}
\begin{center}
    \leavevmode
    \epsfysize=8.0cm
    \epsffile[75 160 575 530]{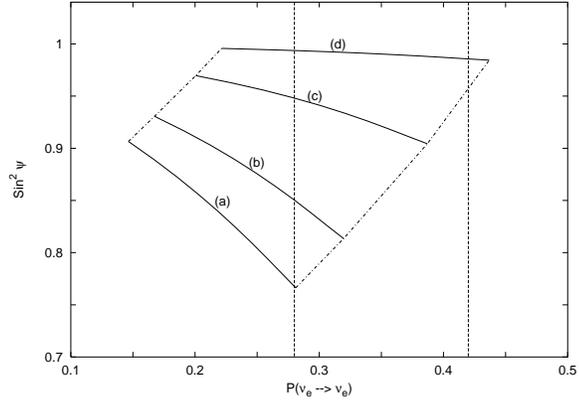}
    \vspace{5.0cm}
\smallskip
 \caption{\label{f2} Plot of $P(\nu_e\rightarrow \nu_e)$
$vs.$ $\sin^2 \psi$ for the ranges of $\tan^2 \theta$ and $\Delta
m^2$ given in Eq. (\ref{kam}) and the same fixed values of
$P(\nu_e\rightarrow \bar\nu_e)$ as in Fig. 1. The two vertical
lines are also  the same as in Fig. 1.}
\end{center}
\end{figure}

\end{document}